\documentclass[aps,prl,reprint,superscriptaddress]{revtex4-2}
\usepackage{lipsum}
\usepackage{graphicx}
\usepackage{amsmath}
\usepackage[colorlinks=true, linkcolor=blue, citecolor=blue, urlcolor=blue]{hyperref}

\begin{document}
\title{Enhanced Excited State Population and Coherence via Adiabatic Tunneling Ionization and Excitation}
\author{Chi-Hong Yuen}
\email[]{cyuen2@kennesaw.edu}
\affiliation{Department of Physics, Kennesaw State University, Marietta, GA 30060, USA}


\begin{abstract}
Tunneling ionization followed by strong-field excitation leads to important ultrafast phenomena such as charge migration and lasing. Recent theoretical developments suggest that the population of the ionic excited state can be greatly enhanced due to the complex interplay between tunneling and excitation. In this Letter, using an adiabatic approach for both tunneling and excitation, semi-analytical solutions are derived for the population and coherence of a two-level ionic system. This approach removes the strong-field dressing, revealing novel sub-half-cycle processes for excited state population and coherence buildup. It predicts that the excited state population is enhanced by an order of magnitude, independent of the laser wavelength, while coherence amplitude can be boosted by over four orders of magnitude for a multi-cycle pulse. For a single-cycle pulse, it suggests that coherence amplitude decreases rapidly as the wavelength increases. This work introduces a novel framework for generating and controlling the electronic excited state and coherence using intense laser pulses, with applications in strong-field control of chemistry and lasing.
\end{abstract}

\maketitle

Tunneling ionization (TI) and strong-field excitation have been known as two fundamental but separated processes in intense laser-matter interactions.
Under an intense low-frequency field, a valence electron in an atom or molecule could tunnel out from the adiabatic potential barrier created by the Coulomb potential and laser field~\cite{Ammosov1986, Tong2002, Tolstikhin2011}, giving rise to a plethora of strong-field phenomena.
The dipole coupling from the intense field could drive multiphoton absorption to form highly excited bound states~\cite{Li2014, Chetty2022, Toth2023}.
Excitation before TI could enhance~\cite{Klaiber2015, Serebryannikov2016} or frustrate~\cite{Nubbemeyer2008, Li2014} the total TI yield.
On the other hand, excitation after TI has received significant attention, as it may create a superposition of ionic states~\cite{Smirnova2009, Kraus2015, Kobayashi2020, Kobayashi2020a, He2022, He2023}, driving electron motion within a molecule.
It could also create ionic coherence for lasing~\cite{Liu2015, Xu2015, Lei2022, Chen2024}, transforming air as a gain medium for propagating ultraviolet light or providing new light sources in the ultraviolet domain.

Most theoretical approaches model excitation after TI in two steps: first, calculate the ionization amplitude or probability to different ionic states, then solve the time-dependent Schr\"{o}dinger equation for the excitation~\cite{Smirnova2009, Kraus2015, Xu2015, He2022, He2023}.
When the ionized electron is neglected, the ion becomes an open quantum system, and a density matrix (DM) formalism should be used~\cite{Rohringer2009, Pfeiffer2013}.
A DM approach has been recently introduced by Zhang \textit{et al.}~\cite{Zhang2020} to describe TI and excitation concurrently.
Yuen and co-workers employed a DM approach for sequential double ionization~\cite{Yuen2022, Yuen2023, Jia2024} and triple ionization of molecules~\cite{Jia2025}.
Yuen and Lin also improved it by accounting for the ionic coherence created by TI~\cite{Yuen2023b, Yuen2024b}.
These DM approaches demonstrate that the population of the excited ionic state, which is far detuned from the laser frequency, is significantly enhanced by strong-field excitation from the ionic ground state formed by TI.
In addition, Chen \textit{et al.}~\cite{Chen2024} recently showed that the ionic coherence created by both TI and excitation at a particular wavelength could further enhance subsequent excitation.
These enhancements, resulting from the complex interplay between TI and excitation, should be applicable to various systems.
They could have significant implications for creating and controlling electronic excited states and coherence in contexts for chemical reaction control~\cite{Brif2010, Lepine2014, Calegari2023} and lasing~\cite{Lei2022}.
Therefore, one is motivated to develop a simple physical model to elucidate the underlying mechanisms.

\begin{figure}[ht]
\includegraphics[width=3.5cm]{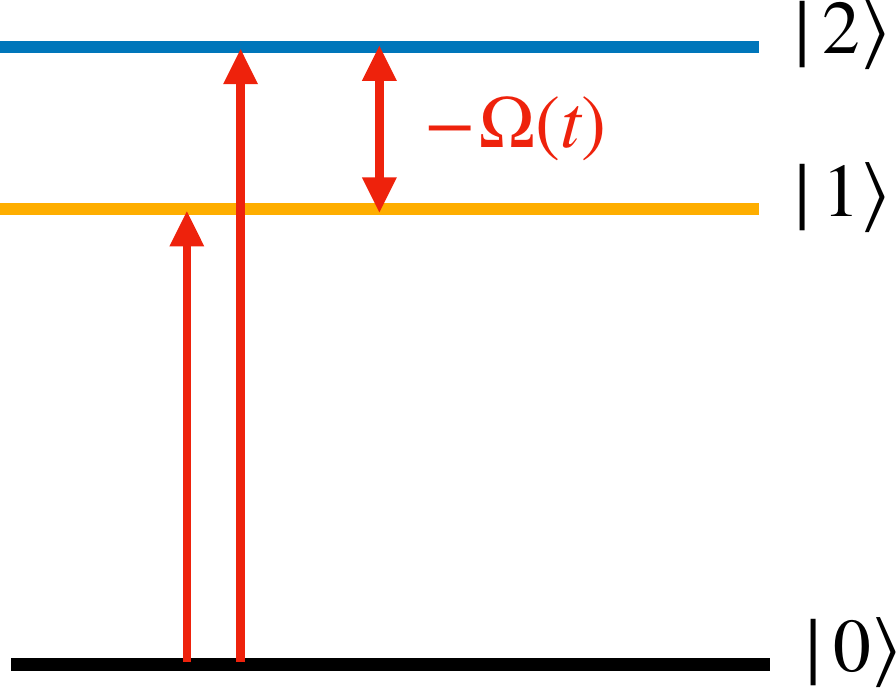}
\caption{Level diagram for a quantum system consisting of a neutral ground state $|0\rangle$ and two ionic states $|1\rangle$ and $|2\rangle$. The $|0\rangle$ state could be tunnel ionized to $|1\rangle$ or $|2\rangle$, which are coupled by $-\Omega(t)$.}
\label{fig:level}
\end{figure}

In this Letter, TI and excitation for a quantum system with a minimal level structure in Fig.~\ref{fig:level} is considered.
Under an intense, low-frequency laser field, the neutral ground state $| 0\rangle$ could be tunnel-ionized adiabatically to dipole-coupled ionic states $|1\rangle$ or $|2\rangle$.
The laser field is treated classically, and multiphoton transitions are not considered.
By treating the excitation adiabatically as well, semi-analytical solutions for the ionic density matrix are derived.
The model demonstrates that adiabatic TI and adiabatic excitation occur simultaneously, unveiling novel sub-half-cycle processes for excited electronic state population and coherence buildup.
It predicts that the population is enhanced by an order of magnitude, independent of laser wavelength, and the coherence amplitude could be enhanced by over four orders of magnitude under a phase-matching condition for the wavelength of a multi-cycle pulse.
The model further suggests that increasing the wavelength for single-cycle pulses will reduce the coherence amplitude.

The derivation begins from the DM approach for strong-field ionization (DM-SFI)~\cite{Yuen2023b, Yuen2024b}. Atomic units are used throughout.
The DM-SFI model numerically solves the equation of motion for the ionic density matrix in the diabatic basis,
\begin{align}
\dot{\rho}^{d}(t) &= -i [H^{d}(t), \rho^{d}(t)] + \Gamma^{d}(t),
\label{eq:DMSFI}
\end{align}
with
\begin{align}
H^{d}(t)
&= \begin{pmatrix}
-\Delta/2 & -\Omega(t) \\
-\Omega(t) &  \Delta/2
\end{pmatrix}
\\
\Gamma^{d}_{ij}(t) &=  \rho_{0}(t) \sum_{m} \gamma_{im}(t) \gamma_{jm}^{\ast}(t).
\end{align}
In the above, $\Delta$ is the energy separation between states $|2\rangle$ and $|1\rangle$, $\Omega = \vec{d}\cdot \vec{F}$ is the dipole coupling, and $\vec{d}$ and $\vec{F}$ is the transition dipole moment and electric field, respectively. $\gamma_{im}$ is the adiabatic ionization amplitude to form the $i$th state with a magnetic quantum number $m$~\cite{Tolstikhin2011, Yuen2023b, Yuen2024b} and $ \rho_{0}(t) = \exp{[-\int_{t_{0}}^{t} \sum_{i,m} |\gamma_{im}(t')|^{2}} dt']$ is the population of state $|0\rangle$.
The initial conditions are $ \rho_{0} (t_{0})=1$ and $ \rho^{d}(t_{0}) = 0$.

The diabatic Hamiltonian can be diagonalized by a mixing matrix $M$,
\begin{align}
M(t)
&= \begin{pmatrix}
\cos \theta(t) & -\sin \theta(t) \\
\sin \theta(t) &  \cos \theta(t)
\end{pmatrix},
\end{align}
with $\tan 2\theta (t) = 2\Omega(t) / \Delta$, and the adiabatic Hamiltonian $H^{a} = M^{T} H^{d} M$ has eigenvalues $\pm E(t) \equiv  \pm \sqrt{ (\Delta/2)^{2} + \Omega^{2}(t)} $.

The equation of motion for the ionic density matrix in the adiabatic basis is
\begin{align}
\dot{\rho}^{a}(t) &= -i [H^{a}(t), \rho^{a}(t)] + \Gamma^{a}(t)  \nonumber \\
& + \dot{M}^{T} \rho^{d} M(t) + M^{T} \rho^{d} \dot{M}(t),
\label{eq:ada}
\end{align}
where $\rho^{a} = M^{T} \rho^{d} M$ and $\Gamma^{a} = M^{T} \Gamma^{d} M$.
Note that all the matrices are hermitian.
The last two terms in Eq.~\eqref{eq:ada} arise from a rotating frame and lead to the mixing of the two adiabatic states.
Therefore, they are referred to as the Coriolis-dipole coupling here.

The key to the derivation is that the effect of Coriolis-dipole coupling can be neglected when the laser frequency $\omega$ is much smaller than $\Delta$, and the effect from the coupling is weak compared to that from $\Gamma^{a}$.
This introduces a second adiabaticity parameter $\gamma_e \equiv \omega/\Delta$ in addition to the Keldysh parameter~\cite{Keldysh1965} $\gamma = \omega \sqrt{2I_p}/F_0$.
When $\gamma_e \ll 1$, the mixing angle is quasistatic. This is the quasistatic approximation (QSA) for strong-field excitation.
Consequently, semi-analytical solutions for $\rho^{a}$,
\begin{align}
\rho^{a}_{ii} (t) &\approx \int_{t_{0}}^{t} \Gamma^{a}_{ii}(t') \, dt', \label{Eq:ada-pop} \\
\rho^{a}_{21} (t) &\approx e^{-2i \int_{t_0}^{t} E(t'') \, dt'' } \int_{t_{0}}^{t} \Gamma^{a}_{21}(t') e^{2i \int_{t_0}^{t'} E(t'') \, dt'' } \, dt',
\label{Eq:ada-coh}
\end{align}
are obtained, and the corresponding diabatic density matrix is
\begin{align}
\tilde{\rho}^d(t) = M(t) \rho^{a}(t) M^{T}(t).
\label{Eq:dia2ada}
\end{align}
As the laser pulse ends, $M = I$ and $\rho^{a}$ coincides with $\tilde{\rho}^d$.
Details of the derivation can be found in Sec. S1 of the Supplemental Material (SM)~\cite{SM}.

Equations \eqref{Eq:ada-pop}--\eqref{Eq:dia2ada} apply to any system with the level structure in Fig.~\ref{fig:level}.
In this Letter, N$_{2}$ molecule aligned with the laser polarization is chosen to be the model system due to the recent interest in air lasing~\cite{Liu2013, Liu2015, Xu2015, Zhang2020, Chen2024} and its simple electronic structure.
The $3\sigma_{g}$ and $2\sigma_{u}$ orbitals of N$_{2}$ has a binding energy of 15.6 and 18.8 eV.
Tunneling ionization of these orbitals~\cite{Zhao2010} forms the $X^{2}\Sigma_{g}$ and $B^{2}\Sigma_{u}$ states of N$_{2}^{+}$ with $\Delta = 3.2$ eV. The two states are coupled by a transition dipole moment $\vec{d} = 0.75 \hat{z}$ $ea_{0}$~\cite{Langhoff1988}.
The laser field is assumed to be a Gaussian pulse linearly polarized in the $z$-direction.
The predictions in this Letter can be verified experimentally by first aligning the molecule, then pumping it as later described, and probing it using attosecond transient absorption spectroscopy~\cite{Kleine2022, Yuen2024c, Zhao2025} or dissociative sequential double ionization spectroscopy~\cite{Yuen2024a}.

\begin{figure}[t]
\centering
\begin{tabular}{c}
\includegraphics[width=8cm]{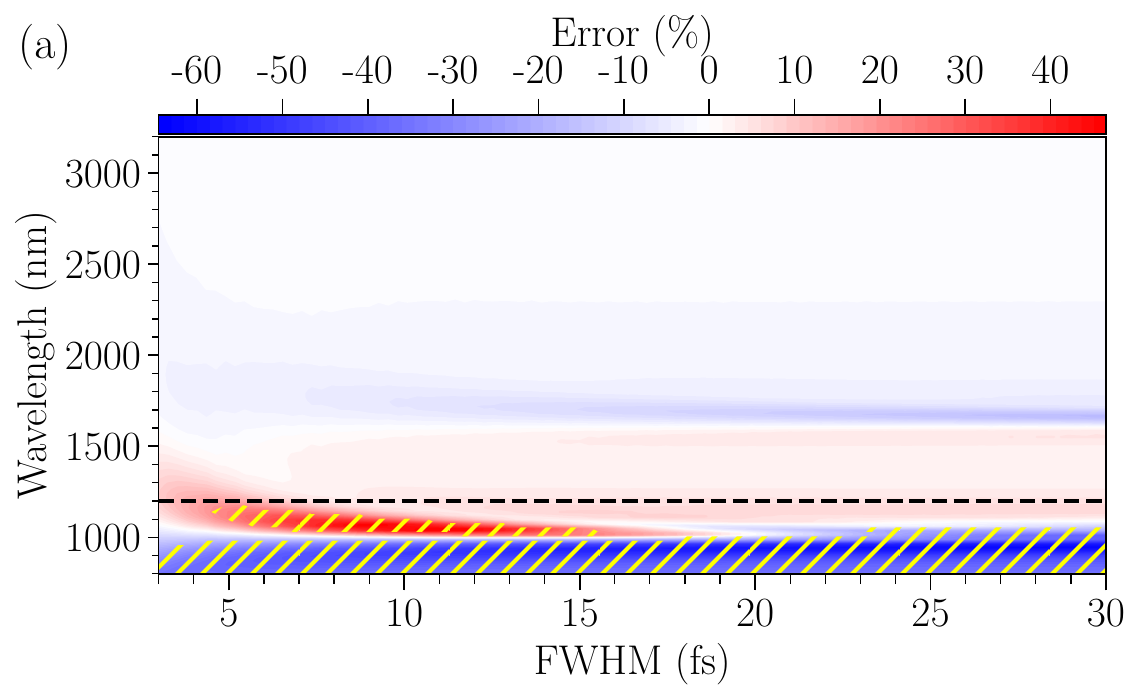} \\
\includegraphics[width=8cm]{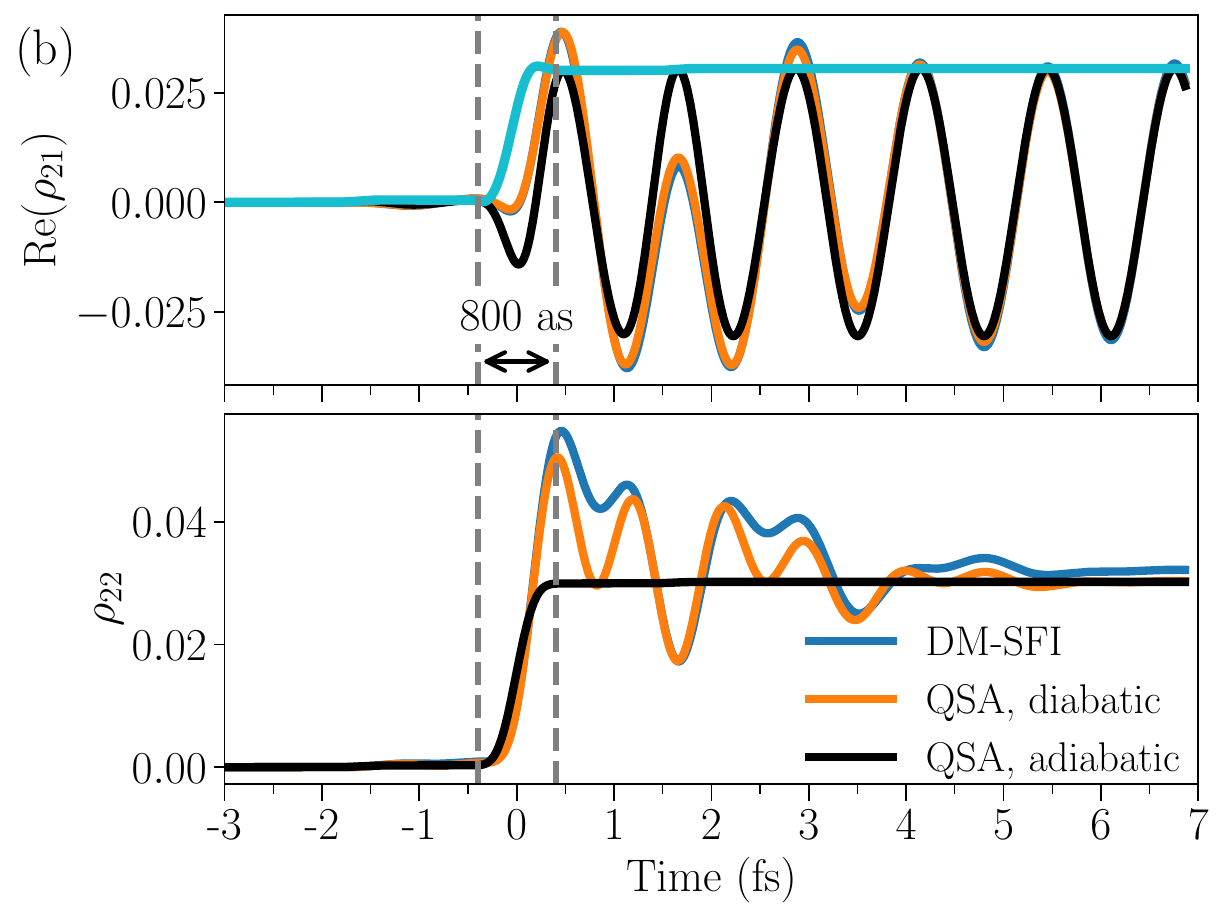}
\end{tabular}
\caption{(a) Error in the excited state population from the quasistatic approximation as a function of wavelength and pulse duration, with a fixed peak intensity of $2 \times 10^{14}$ W/cm$^{2}$.
The hatches mark the region with errors larger than 20\%.
(b) Excited state population and the real part of coherence for a single-cycle 1030 nm laser, calculated using the DM-SFI model (blue line) and the semi-analytical formula in the diabatic (orange line) and adiabatic (black line) basis.
The cyan curve is the amplitude of the adiabatic coherence.}
\label{fig:QSA}
\end{figure}

Before studying the semi-analytical solutions from Eqs. \eqref{Eq:ada-pop} and \eqref{Eq:ada-coh}, one must first validate the QSA.
The errors associated with the QSA can be quantified by difference in the final excited state population, $[\rho^{a}_{22}(t_{f}) - \rho^{d}_{22}(t_{f})]/\rho^{d}_{22}(t_{f}) \times 100\%$, where $\rho^{d}_{22}$ and $\rho^{a}_{22}$ is calculated using the DM-SFI model and semi-analytical formula, respectively.
Figure~\ref{fig:QSA}a displays the errors as a function of wavelength and pulse duration at a fixed peak intensity of $2 \times 10^{14}$ W/cm$^{2}$.
This laser intensity is chosen to avoid completely depleting the neutral state before the pulse ends for the entire range of pulse durations.
It is clear that the error from QSA largely depends on wavelengths. For wavelengths above 1200 nm ($\gamma_e < 0.33$), the error is less than 20 \%. 
For wavelengths below 1200 nm, the error from QSA decreases when the pulse duration is near a single optical cycle because the effect of the Coriolis-dipole coupling is weaker than the TI.
Notably, the error from QSA is less than 10\% for a single-cycle 1030 nm laser ($\gamma_e =0.38$).
This case is particularly interesting because such a pulse has been generated by several experimental groups~\cite{Tsai2022, Han2025, Pi2025}.
Further discussion on the QSA error can be found in Sec. S2 of \cite{SM}.

Figure~\ref{fig:QSA}b shows the evolution of coherence and excited state population for a single-cycle 1030 nm laser.
The semi-analytical solution $\rho^{a}(t)$ as well as its diabatic counterpart $\tilde{\rho}^d(t)$ are plotted with $\rho^d(t)$ from the DM-SFI model.
Only the real part of the coherence is presented here. A comparison of its imaginary part is discussed in detail in Sec. S2 of \cite{SM}.
As the pulse ends at around $t = 6$ fs, the semi-analytical solutions agree well with the DM-SFI model for both the population and coherence.
The real part of $\tilde{\rho}^d_{21}$ is almost indistinguishable from $\rho^d_{21}$.
Between $t=0.5$ and 3.5 fs, the diabatic coherence amplitude changes while the adiabatic coherence amplitude remains constant.
According to Eq.~\eqref{Eq:dia2ada}, such changes result from the dressing of the laser field and do not contribute to net coherence buildup.
By taking the absolute value of the adiabatic coherence, it is clear that the coherence buildup only takes around 800 as near the peak of the pulse.

On the other hand, the semi-analytical diabatic population $\tilde{\rho}^d_{22}$ agrees well with $\rho^{d}_{22}$ from the DM-SFI model, except $\tilde{\rho}^d_{22}$ drops more than $\rho^{d}_{22}$ at around $t=$1 and 2.5 fs.
Such disagreement can be attributed to the Coriolis-dipole coupling and leads to only a 6\% difference in the population.
Therefore, the oscillations in the diabatic population are the effect of strong field dressing and do not contribute to the net population buildup.
As in the case of coherence, the excited state population only takes around 800 as.

To elucidate the mechanisms of the sub-half-cycle buildup of coherence and excited state population, the weak coupling limit $[2\Omega(t) / \Delta]^{2} \ll 1$ is considered.
Full derivation of Eqs.~\eqref{eq:gam22} -- \eqref{Eq:weak-coh} can be found in Sec. S3 of \cite{SM}.
Keeping up to the second-order terms of $2\Omega(t) / \Delta$, one has
\begin{align}
\Gamma^{a}_{22} &= \Gamma^{d}_{11} \sin^{2}\theta + \Gamma^{d}_{22} \cos^{2}\theta - \mathrm{Re}[\Gamma^{d}_{21}] \sin 2\theta \nonumber \\
&\approx \Gamma^{d}_{11} \frac{\Omega^{2}(t)}{\Delta^{2}} + \Gamma^{d}_{22} ( 1-\frac{\Omega^{2}(t)}{\Delta^{2}}) - 2 \mathrm{Re}[\Gamma^{d}_{21}] \frac{\Omega(t)}{\Delta}.  \label{eq:gam22} 
\end{align}

The first term has a physical meaning of shake-up TI, co-occurring with the direct TI to the ionic ground state.
It scales with the square of the dipole coupling so that it builds up the population every half-cycle.
The second term is the direct TI to the excited state and the shake-down TI, which is a weak process since $\Gamma^{d}_{22}$ is much smaller than $\Gamma^{d}_{11}$.
The last term implies that TI coherence (TIC)~\cite{Yuen2023b} induces population transfer in the presence of dipole coupling. This term will be referred to as the TIC-induced transfer.
It is emphasized that all three terms do not explicitly depend on laser wavelength.
As a result, even for an intense mid-infrared laser where $\Delta$ could be far detuned with the photon energy, the excited state could be efficiently populated.

The contribution of each term to the population in the strong field case can be evaluated using Eq.~\eqref{Eq:ada-pop}.
For the case in Fig.~\ref{fig:QSA}b ($[2\Omega(0) / \Delta]^{2} = 0.94$), the term related to shake-up TI, direct TI, and TIC-induced transfer contributes 47\%, 10\%, and 42\% to the total population, respectively.
It indicates that shake-up TI and TIC-induced transfer increases the population by an order of magnitude compared to direct TI, and that TIC should not be neglected in the strong-field case.

Regarding the coherence buildup, one has
\begin{align}
\Gamma^{a}_{21} &=  (\Gamma^{d}_{22} - \Gamma^{d}_{11}) \sin 2\theta/2 + (\Gamma^{d}_{21} \cos^{2}\theta - \Gamma^{d}_{12} \sin^{2}\theta) \nonumber \\
&\approx  (\Gamma^{d}_{22} - \Gamma^{d}_{11}) \Omega(t)/\Delta  + (\Gamma^{d}_{21} - 2 \mathrm{Re}[\Gamma^{d}_{21}] \Omega^{2}(t)/\Delta^{2}). \label{Eq:gam21}
\end{align}
The first term, driven by TI and dipole coupling, changes sign every half-cycle.
The second term is driven by TIC and dipole coupling.
The $3\sigma_{g}$ and $2\sigma_{u}$ orbitals of N$_{2}$ have opposite parity, so that TIC also changes sign every half-cycle~\cite{Yuen2023b, Yuen2024b}.
For the case in Fig.~\ref{fig:QSA}b, direct integration of the terms according to Eq.~\eqref{Eq:ada-coh} shows that the magnitude of the TI-driven term is about 2.4 times larger than the TIC-driven term.

The coherence buildup is more complicated to analyze than the population due to the dynamic phase of the integrand in Eq.~\eqref{Eq:ada-coh}.
However, in the weak coupling limit, one can show that, for a long pulse, the coherence after the pulse ends is
\begin{align}
\rho^{a}_{21} (t) &\approx e^{-i \Delta t} \int_{t_{0}}^{t} \Gamma^{a}_{21}(t') e^{i [(\Delta + \alpha)t' - \eta_{0}] } \, dt',
\label{Eq:weak-coh}
\end{align}
where $\alpha = d^{2}F_{0}^{2}/\Delta$ is the peak ac Stark shift, $\eta_{0} = (\alpha \tau/4) \sqrt{\pi/\ln 2}$ is an overall phase shift, $F_{0}$ is the peak field strength, and $\tau$ is the pulse duration.
Equations~\eqref{Eq:gam21} and \eqref{Eq:weak-coh} imply that if the time-averaged energy separation for adiabatic states $\Delta + \alpha = (2n+1) \omega$ for integer $n$, then the phase changes sign every half-cycle, and the coherence adds up constructively.

\begin{figure}[ht]
\centering
\begin{tabular}{c}
\includegraphics[width=8cm]{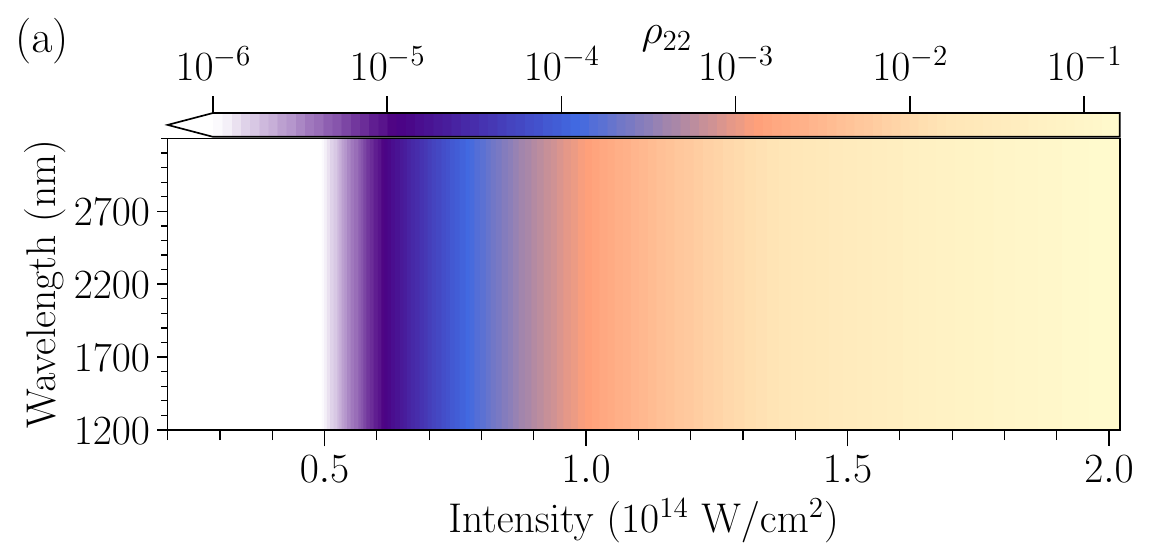} \\
\includegraphics[width=8cm]{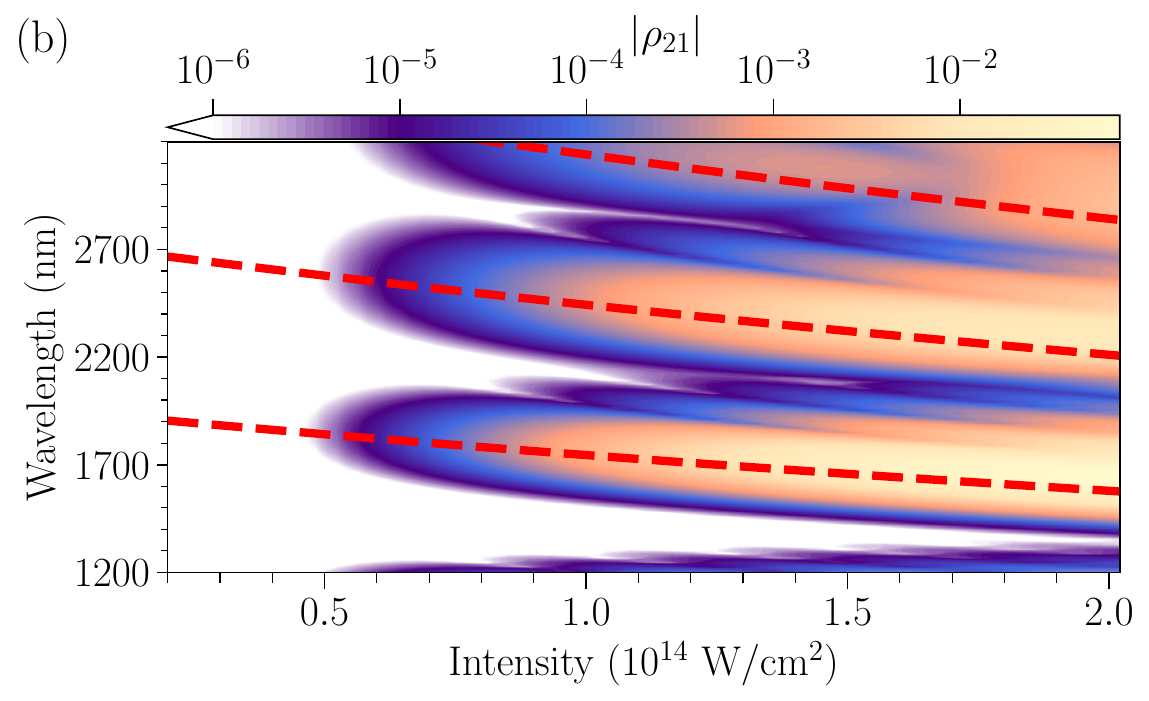}
\end{tabular}
\caption{
(a) Excited state population calculated using the two-level model for a 30 fs pulse at different wavelengths and peak intensities.
(b) Same as (a), but for the amplitude of coherence. The dashed lines trace out the phase-matching condition for five, seven, and nine photons at the weak-coupling limit.
}
\label{fig:2Dscan}
\end{figure}

To verify the buildup mechanisms, excited state population and coherence amplitude are calculated using Eqs.~\eqref{Eq:ada-pop} and \eqref{Eq:ada-coh} for a 30 fs pulse at wavelengths from 1200 to 3200 nm and at intensities from 0.2 to 2 $\times 10^{14}$ W/cm$^{2}$.
These parameter ranges were chosen because the error from QSA is negligible (see Fig. S1 of \cite{SM}).
Figure~\ref{fig:2Dscan}a demonstrates that the excited state population solely depends on the peak laser intensity and is indeed an adiabatic process, i.e., independent of wavelength.
Figure~\ref{fig:2Dscan}b plots the coherence amplitude after the pulse ended.
The maximum amplitudes closely follow the phase-matching conditions for five, seven, and nine photons, even in the strong coupling regime.
At 2 $\times 10^{14}$ W/cm$^{2}$, the phase-matched wavelength is around 1644 nm, corresponding to $\Delta + \alpha = 5.2 \omega$.
At the same peak intensity, the coherence amplitude at around 1341 nm, corresponding to $\Delta + \alpha = 4.2 \omega$, is dramatically suppressed by more than four orders of magnitude.
The above results showcase that, under the phase-matching condition, controlling the wavelength can significantly enhance the coherence amplitude for a multi-cycle pulse.

\begin{figure}[ht]
\centering
\includegraphics[width=7cm]{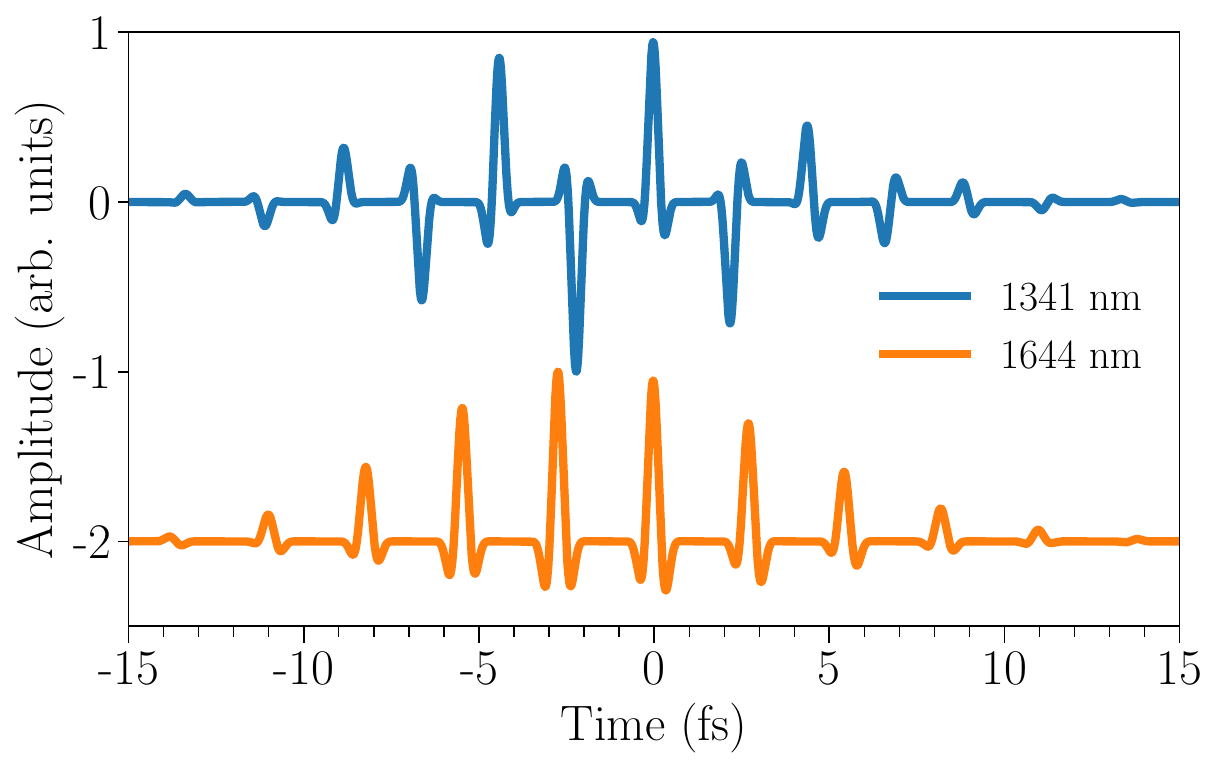}
\caption{
Real part of the buildup function of the adiabatic coherence when the final coherence amplitude is at its minimum (blue line) and maximum (orange line) for a 30 fs pulse with a peak intensity of $2 \times 10^{14}$ W/cm$^{2}$.
}
\label{fig:buildup}
\end{figure}

To see the buildup of coherence amplitude in the phase-mismatched and phase-matched scenarios, Fig.~\ref{fig:buildup} plots the real part of the integrand in Eq.~\eqref{Eq:ada-coh} for the 1341 and 1644 nm laser with 30 fs pulse duration and $2 \times 10^{14}$ W/cm$^{2}$ peak intensity.
For the 1341 nm laser, the peak values of the buildup function change sign every half-cycle, so the coherence adds up destructively.
In contrast, the peak values of the buildup function for the 1644 nm laser have the same sign every half-cycle, resulting in coherence that adds up constructively.
Note that the lineshapes are not identical every half-cycle as the dynamics phase in the strong coupling case is more complicated than in Eq.~\eqref{Eq:weak-coh}.

\begin{figure}[t]
\centering
\begin{tabular}{@{\hskip 0pt}c@{\hskip 0mm}c@{\hskip 0pt}}
\includegraphics[width=4.3cm]{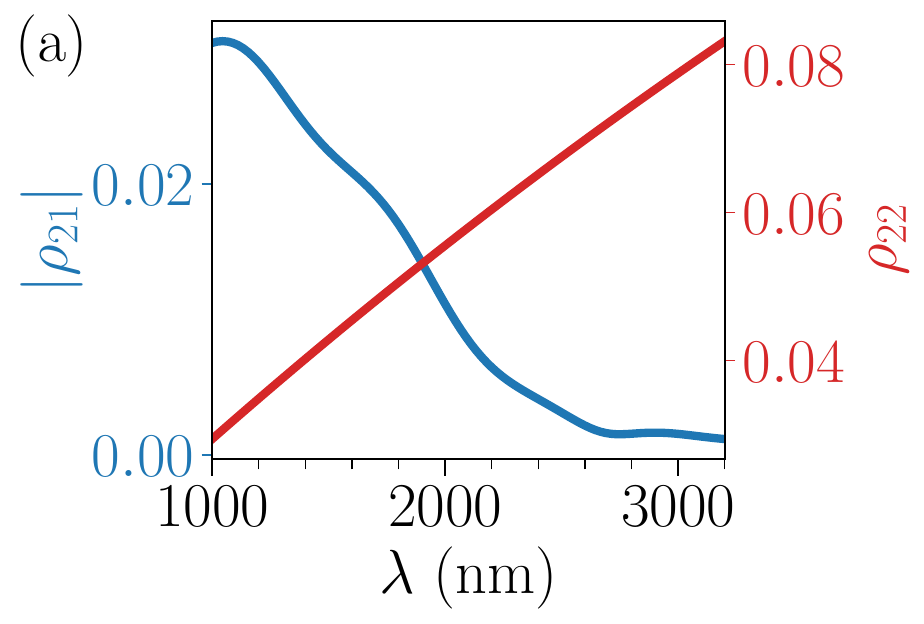} &
\includegraphics[width=4.3cm]{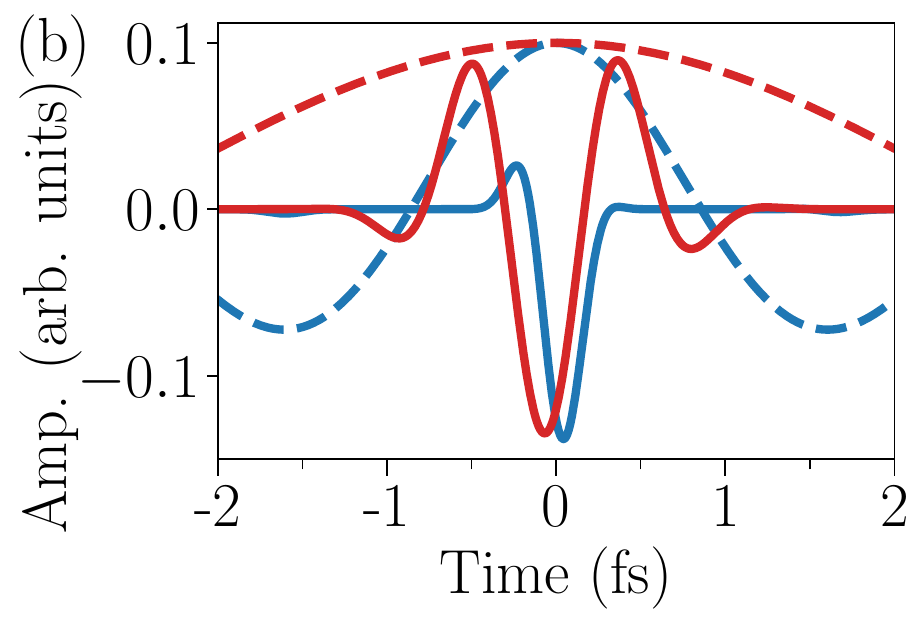}
\end{tabular}
\caption{(a) Amplitude of coherence (blue line) and excited state population (red line) generated by a single-cycle laser pulse with a peak intensity of $2 \times 10^{14}$ W/cm$^{2}$ at different wavelengths.
(b) Real part of the buildup function of the adiabatic coherence for the 1030 (blue solid line) and 3200 nm (red solid line) laser. The dashed lines of the same color are the corresponding electric field profiles.}
\label{fig:1cycle}
\end{figure}

The recent development of sub- and single-cycle pulses from near-infrared~\cite{Rossi2020, Tsai2022, Steinleitner2022, Kowalczyk2023, Han2025, Pi2025} to mid-infrared~\cite{Liang2017} raises the question of whether the coherence amplitude can also be significantly enhanced by changing the wavelength of a single-cycle pulse.
Figure~\ref{fig:1cycle}a shows the coherence amplitude and excited state population created by a single-cycle pulse with a peak intensity of $2 \times 10^{14}$ W/cm$^{2}$ at different wavelengths.
Although the population increases with pulse duration and wavelength, the coherence amplitude decreases rapidly as the wavelength increases.
The coherence amplitude of the 1030 nm ($\tau = 3.4$ fs) laser is about 26 times greater than that of the 3200 nm laser ($\tau = 10.7$ fs).
This can be explained by the dynamics phase in the coherence buildup.
Figure~\ref{fig:1cycle}b plots the real part of the integrand in Eq.~\eqref{Eq:ada-coh} for the 1030 and 3200 nm lasers along with their electric field profiles.
For the 3200 nm laser, the broader pulse envelope leads to a uniform dynamics phase, resulting in a symmetric buildup function that cancels out coherence.
In contrast, for the 1030 nm laser, the narrower pulse envelope results in a highly asymmetrical buildup function, leading to net coherence buildup.
As a result, using a single-cycle pulse with a shorter wavelength is advantageous for enhancing the coherence amplitude.

Finally, the intensity control of the coherence amplitude for a single-cycle laser is trivial, as the amplitude scales with the TI rate.
Carrier-envelope phase (CEP) control is found to be inefficient as the amplitude changes less than 10\%.
Interestingly, if $\Delta + \alpha \approx (2n+1) \omega$, the phase of the coherence will change $(2n+1)$ times the CEP as the CEP varies from 0 to $2\pi$.
See Sec. S4 of \cite{SM} for detailed discussions.

To summarize, accurate semi-analytical solutions for a two-level ionic density matrix are obtained using a novel adiabatic theory for tunneling ionization and strong-field excitation.
The approach removes the strong-field dressing and reveals novel sub-half-cycle processes that significantly enhance excited state population and coherence buildup.
These new findings offer a novel framework for future research opportunities in creating and controlling excited states or electronic coherence in molecules using intense laser fields, with applications in strong-field control of chemistry and lasing.

\textit{Acknowledgements} ---
The author thanks C.D. Lin, Chuan Cheng, and Meng Han for valuable comments.
This work was supported by the start-up fund provided by the College of Science and Mathematics at Kennesaw State University.


%

\end{document}